\documentclass[preprint]{aa}
\usepackage[varg]{txfonts}
\usepackage{multirow}
\usepackage{natbib}
\usepackage{graphicx}

\begin{document}

\title{Orbital evolution of a tidally stripped star and disk-driven stable mass transfer \\ for QPEs in GSN 069}
\author{Di Wang}
\institute{Department of Physics, Huazhong University of Science and Technology, Wuhan 430074, China}
\abstract
{ A plausible origin for the quasi-periodic eruptions (QPEs) could be mass loss at the periastron of a body moving around the supermassive black hole (SMBH) on a highly eccentric orbit. This kind of tidally stripped star is expected to radiate gravitational waves, thereby leading to a shrinkage of the periastron distance. As a result, it will eventually be disrupted by the SMBH, as suggested by previous studies. }
{This scenario predicts a gradually increasing mass transfer, contradicting the long-term evolution of the observed intensity of QPEs in GSN 069.}
{
    In this work, we revisit the orbital evolution of the stripped star and we propose a model of a tidally stripped WD moving inside an accretion disk for QPEs, aimed at characterizing GSN 069.
}
{We found the effect of the mass transfer ultimately dominates the orbital evolution, resulting in the stripped star finally escaping the SMBH, rather than being disrupted by it. The drag force induced by the disk may effectively reduce the mass transfer and could thus explain the observed long-term evolution in the intensity of the QPEs in GSN 069. The disk is likely a fallback disk of the tidal disruption event in GSN 069. Considering the evolution of its accretion rate, this scenario could also explain  the increase in the intensity of the latest eruption.
}
{}
\maketitle
\titlerunning{Orbital evolution of the tidally stripped star}
\section{Introduction}
A white dwarf (WD) orbiting around a supermassive black hole (SMBH) with a high eccentricity may lose mass each time via the periastron. Such a tidally stripped WD is not only expected to emit periodic electromagnetic wave signals \citep{zalamea2010white}, but it is also a target for space-based gravitational wave detectors, such as Laser Interferometer Space Antenna (LISA, \citealt{babak2017science}) and TianQin \citep{luo2016tianqin}. The first discovered quasi-periodic eruptions (QPEs) from GSN 069 are suggested to have possibly originated from a tidally stripped WD on a highly eccentric orbit \citep{miniutti2019nine,wang2022model,king2020gsn}.

Due to gravitational wave radiation decreasing the periastron distance, the mass loss of the tidally stripped WD gradually increases until, finally, the WD is tidally disrupted by the SMBH \citep{zalamea2010white,wang2022model}. However, the intensity of QPEs in GSN 069 shows no sign of increasing \citep{miniutti2023repeating,miniutti2023alive}, which is inconsistent with the cited prediction. \cite{king2022quasi} pointed out that the mass transfer stability requires the angular momentum transfer from the disk to the orbiting star. He suggested that an orbital resonance with the disk possibly could accomplish this, but any further discussion is still lacking.

GSN 069 also has a tidal disruption event (TDE) that has lasted more than a decade \citep{shu2018long,sheng2021evidence}. This TDE was recently found to have re-brightened and, while it was brightening, the QPEs disappeared \citep{miniutti2023repeating}. When radiation from the TDE faded again, the QPEs reappeared, and one of two detected QPEs was stronger than in the last outburst before they disappeared \citep{miniutti2023alive}. Such a behavior implies that the evolution of mass transfer may be related to the accretion rate of the accretion disk.

The origin of this TDE is likely caused by the disruption of a common envelope of the binary containing the current orbiting body before it is captured by the SMBH \citep{wang2024tidal}. This means that the fallback disk of the TDE is nearly coplanar with the orbit of the orbiting body that produced the QPEs. In this situation, the drag force by the disk may significantly affect the secular orbital evolution of the orbiting body.

In this paper, we first revisit the secular evolution of the tidally stripped star with high eccentricity, taking into account the effect of the mass transfer on the orbit, which has been ignored in the previous studies. We found that the effect of mass transfer significantly changes the criterion for unstable mass transfer. The mass transfer for WDs is no longer dynamically stable, while it remains dynamically stable for main sequence stars. Furthermore, the effect of the mass transfer dominates the secular orbital evolution, which leads to the fate of the stripped star escaping the SMBH, rather than being disrupted by it. 

While the origin of QPEs is still a matter of debate, there are various other models to consider in this context \citep{chen2022milli,zhao2022quasi,wang2022model,krolik2022quasiperiodic,lu2023quasi,xian2021x,sukova2021stellar,linial2023unstable,linial2023emri+,franchini2023quasi}. In this paper, we focus on the stripped WD model on a highly eccentric orbit for GSN 069 \citep{king2020gsn,king2022quasi}. We calculated the secular effect of drag force by the disk on the orbit for GSN 069. We found that the drag force can transfer the angular momentum of the disc to the stripped WD, and then stabilize the mass transfer. This effect is strongly dependent on the accretion rate, which can be estimated from the luminosity of the TDE. The derived change in the mass transfer rate by the disk drag features the ability to suppress other effects that can increase mass transfer, thus explaining the long-term evolution of the observed intensity of QPEs. The increase in the intensity of the partial QPEs after the TDE re-brightened may also be explained by this model.

This paper is organized as follows. In Section \ref{sec2}, we revisit the secular orbital evolution of the stripped star under the effect of gravitational wave radiation and mass transfer. In Section \ref{sec3}, we discuss the effect of drag force by the accretion disk on the orbit for QPEs in GSN 069. We present our discussion in Section \ref{sec4} and a summary in Section \ref{sec5}.

\section{Evolution with mass transfer and gravitational radiation}\label{sec2}
We consider a WD of mass $M$ and radius $R$ on a highly eccentric orbit around a SMBH of mass $M_h$. When it reaches the periastron, the stripped mass can be obtained by calculating the mass of the matter in the WD outside Roche lobe \citep{chen2023tidal}:
\begin{equation}\label{eq: dm}
    \frac{\Delta M}{M} \simeq 4.8\left[1-\left(M / M_{\mathrm{ch}}\right)^{4 / 3}\right]^{3 / 4}\left(1-\frac{\beta_0}{\beta}\right)^{5 / 2}.
\end{equation}
Here, $\beta\equiv r_t/r_p$ is impact factor, where $r_t=R(M_h/M)^{1/3}$ is the disrupted radius and $r_p$ is the periastron radius. The Chandrasekhar mass $M_{\mathrm{ch}}$ is $1.44 M_{\odot}$ and $\beta_0$ is about 0.5. Mass transfer occurs only if $\beta$ is greater than $\beta_0$. And the WD is disrupted when $\beta$ is greater than 1. 

Differentiating it then gives:
\begin{equation}\label{eq: ddm}
    \frac{\dot{\Delta M}}{\Delta M}=\frac{1-2\left(M / M_{\mathrm{ch}}\right)^{4 / 3}}{1-\left(M / M_{\mathrm{ch}}\right)^{4 / 3}}\frac{\dot{M}}{M}+\frac{5\beta_0}{2(\beta-\beta_0)}\frac{\dot{\beta}}{\beta}.
\end{equation}
The latter term is usually dominant because $\beta$ is close to $\beta_0$ when there is not a great amount of mass transfer. Setting the mass-radius index, $\zeta,$ which gives mass-radius relation of $R \propto M^\zeta$, then $\dot{\beta}$ can be obtained via \citep{king2022quasi}:
\begin{equation}\label{eq: beta_dot}
    \frac{\dot{\beta}}{\beta}=\frac{2 \dot{M}}{M}\left(\frac{5}{6}+\frac{\zeta}{2}-q\right)-\frac{2 \dot{J}}{J}+\frac{\dot{e}}{1+e}.
\end{equation}
Here, $q=M/M_h$ is the mass ratio, $J$ is the total orbital angular momentum and $e$ is the orbital eccentricity. 
\subsection{Mass transfer}
The mass-radius relation of the WD is \citep{paczynski1971evolutionary}:
\begin{equation}\label{eq: m_r_relation}
    R=9 \times 10^8\left[1-\left(\frac{M}{M_{\mathrm{ch}}}\right)^{4 / 3}\right]^{1 / 2}\left(\frac{M}{M_{\odot}}\right)^{-1 / 3} \mathrm{~cm},
\end{equation}
which is adopted in the Eq. \eqref{eq: dm} derivation \citep{chen2023tidal}. So, $\zeta$ for the WD is about -1/3 for a low mass and becomes even more negative when mass is close to $M_{\mathrm{ch}}$. Furthermore, it is always greater than -5/3, so that the first term in Eq. \eqref{eq: beta_dot} is negative, which can suppress mass transfer. \cite{king2022quasi} claimed that this term is dominant on short timescales, so the mass transfer is dynamically stable. However, he ignored the $\dot{e}$ caused by the mass transfer, which is\footnote{\cite{wang2022model} considered the effect of the mass transfer on the orbit as well. Their formula is different from this because they mistake the instantaneous mass transfer rate in \cite{sepinsky2007interacting} for the secular mass transfer rate, thus underestimating this effect.} \citep{sepinsky2007interacting}:
\begin{equation}\label{eq: edot_mdot}
    \langle\dot{e}\rangle_{\dot{M}}=-2(1+e)\frac{\dot{M}}{M}(1-q).
\end{equation}
Substituting it into Eq. \eqref{eq: beta_dot} gives
\begin{equation}\label{eq: beta_dot_mdot}
    \left\langle\frac{\dot{\beta}}{\beta}\right\rangle_{\dot{M}}=\frac{\dot{M}}{M}\left(\zeta-\frac{1}{3}\right).
\end{equation}
Here, $\langle\dot{J}/J\rangle\approx 0$, because only very little angular momentum is removed by central SMBH via accretion\citep{king2022quasi}.

So the self-driven mass transfer for the WDs is unstable. Eq. \eqref{eq: edot_mdot} shows the orbit is more eccentric under mass transfer and the change in the semi-major axis \citep{sepinsky2007interacting}:
\begin{equation}\label{eq: adot_mdot}
    \left\langle\frac{\dot{a}}{a}\right\rangle_{\dot{M}}=-2\frac{1+e}{1-e}\frac{\dot{M}}{M}(1-q),
\end{equation}
is positive. So unstable mass transfer would lead to the eventual escape of the WD, contrary to the prediction that gravitational radiation would lead to the disruption of the WD \citep{zalamea2010white,wang2022model}.
\subsection{Gravitational radiation}
The change in orbit caused by the gravitational radiation is \citep{peters1964gravitational}
\begin{equation}\label{eq: Jdot_gw}
    \left\langle\frac{\dot{J}}{J}\right\rangle_{GW}=-\frac{32}{5} \frac{G^{3} M M_hM_t}{c^5 a^{4}\left(1-e^2\right)^{5/2}}\left(1+\frac{7}{8} e^2\right),
\end{equation}
\begin{equation}\label{eq: adot_gw}
    \left\langle\frac{\dot{a}}{a}\right\rangle_{GW}=-\frac{64}{5} \frac{G^{3} M M_h M_t}{c^5 a^4\left(1-e^2\right)^{7 / 2}}\left(1+\frac{73}{24}e^2+\frac{37}{96}e^4\right),
\end{equation}
\begin{equation}\label{eq: edot_gw}
    \langle\dot{e}\rangle_{GW}=-\frac{304}{15} e \frac{G^{3} M M_h M_t}{c^5 a^4\left(1-e^2\right)^{5 / 2}}\left(1+\frac{121}{304} e^2\right).
\end{equation}
Here, $G$ is the gravitational constant, $M_t=M_h+M$ is total mass, and $c$ is the speed of light. Substituting Eq. \eqref{eq: Jdot_gw} and \eqref{eq: edot_gw} into Eq. \eqref{eq: beta_dot} yields
\begin{equation}
    \left\langle\frac{\dot{\beta}}{\beta}\right\rangle_{GW}=\frac{64}{5(1+e)} \frac{G^{3} M M_h M_t}{c^5 a^4\left(1-e^2\right)^{5 / 2}}\left(1-\frac{7}{12}e+\frac{7}{8}e^2+\frac{47}{192}e^3\right),
\end{equation}
which is positive. So both mass transfer and gravitational radiation enhance mass transfer, but gravitational wave radiation reduces $a$ and $e$, which is the opposite of the effect of mass transfer. 
\subsection{Orbital evolution for the WD}
\label{sec: orbital_evolution_for_the_wd}
The effect of the gravitational radiation is stronger than the effect of the mass transfer when the mass transfer rate is low. As the mass transfer rate gets higher, the effect of mass transfer becomes dominant. While factor $1/(1-e)$ evolves to very high values, the effect of gravitational radiation becomes smaller. Because the periastron $r_p=a(1-e)$ is constant under mass transfer, $\langle\dot{r_p}\rangle_{\dot{M}}=0$, this makes the effect of gravitational radiation proportional to $(1-e)^{1/2}$ for $\dot{a}/a$ and $(1-e)^{3/2}$ for $\dot{J}/J$ and $\dot{e}$ when $e\approx 1$.

The mass transfer rate $|\dot{M}|$ doesn't grow endlessly. Because $\dot{M}$ is related not only to $\Delta M$ but also to the orbital period $P_b$, $\dot{M} = -\Delta M/P_b$. And then $a$ grows dramatically at high $e$, as does $P_b$.  When the growth rate of $P_b$ is greater than the growth rate of $\Delta M$, $|\dot{M}|$ will start to decrease until the WD escapes. And the critical condition can be obtained by: 
\begin{equation}\label{eq: dm_balance}
    \frac{\ddot{M}}{\dot{M}}= \frac{\dot{\Delta M}}{\Delta M}-\frac{3}{2}\frac{\dot{a}}{a}\approx \left[\frac{5\beta_0}{2(\beta-\beta_0)}\left(\zeta-\frac{1}{3}\right)+3\frac{1+e}{1-e}\right]\frac{\dot{M}}{M}=0.
\end{equation}

So, the WD finally escapes with eccentricity $e_f\gtrsim  1$, then the total mass it loses can be obtained from Eq. \eqref{eq: edot_mdot}:
\begin{equation}\label{eq: mass_loss_mdot}
    -\frac{\Delta M}{M}\approx \frac{e_f-e_i}{2(1+e_i)(1+q)}\approx \frac{1-e_i}{4(1+e_i)},
\end{equation}
where $q\ll 1$ and $e_i$ is the initial eccentricity while the effect of the mass transfer is dominant. The WD loses at most 1/4 of its total mass and can survive the mass loss.

\begin{figure}
    \centering
    \includegraphics[width=\linewidth]{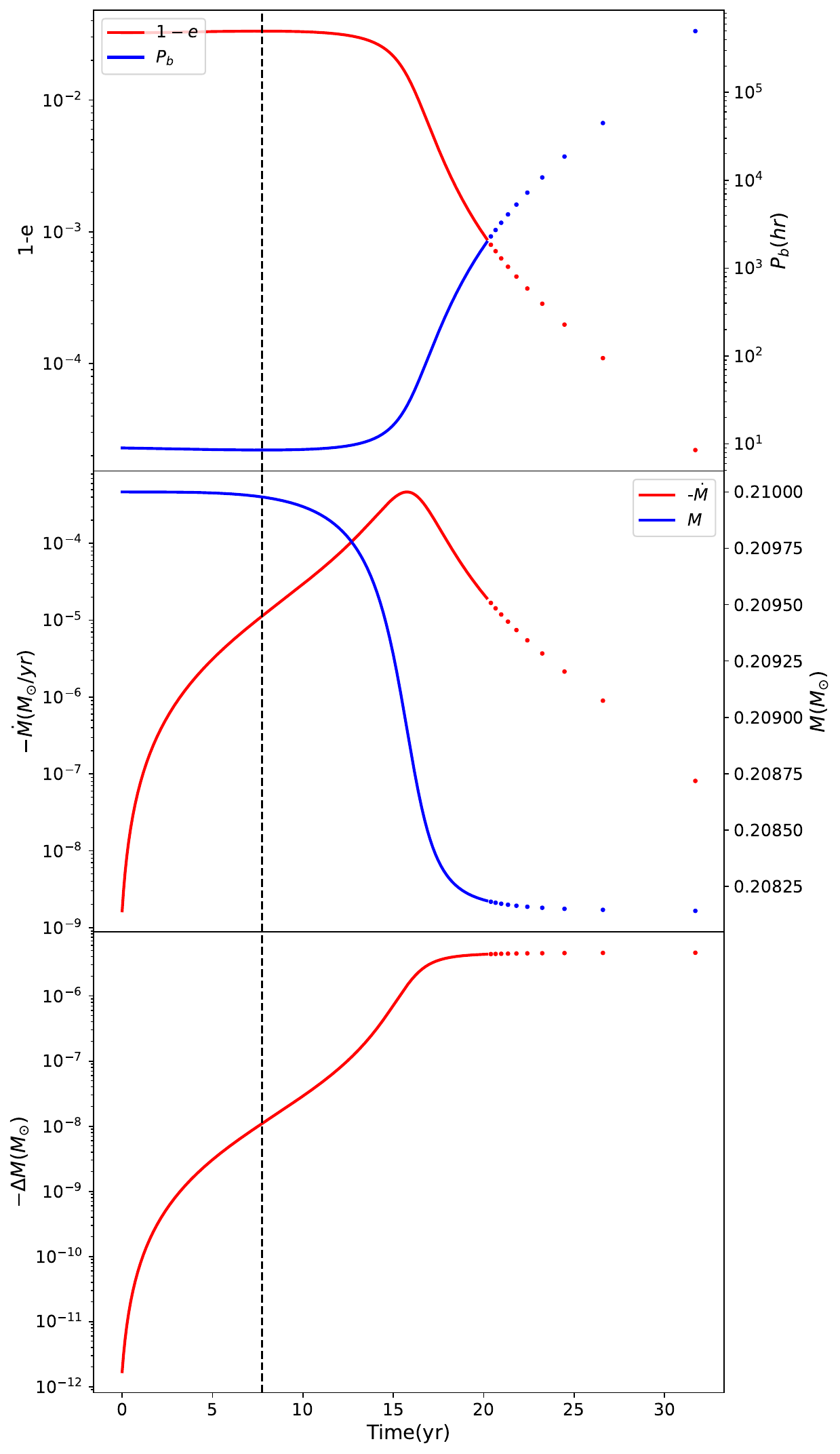}
    \caption{Evolution of orbit parameters and WD's mass. The first panel from top to bottom shows the evolution of $1-e$ and orbital period $P_b$. The second one is the evolution of the average mass transfer rate and mass of the WD. And the last one represents the evolution of mass loss at periastron. The dashed line represents the critical time that the effect of the mass transfer becomes dominant. For each parameter, the critical times are not the same but very close. Here we use the critical time of $e$ for all panels. The last ten orbits are plotted as scatter plots and the rest as line plots.\label{fig: evolution_wd}}
\end{figure}

Here, we show an example of the evolution under mass transfer and gravitational radiation in Fig. \ref{fig: evolution_wd}. We adopted plausible parameters for GSN 069: $P_b=9 \mathrm{~hr}$, $M_h=5\times 10^5 M_{\odot}$, and $M=0.21 M_{\odot}$. Then, $e=0.968$ can be obtained from Eq. \eqref{eq: m_r_relation} with $\beta\approx \beta_0$. The initial $\beta$ is 0.50001. The orbital parameters and the mass and radius of the WD are updated each time the WD passes through the periastron. We use $\langle\dot{r_p}\rangle$ and $\langle\dot{e}\rangle$ to evolve orbit, because $\langle\dot{a}/a\rangle $ is too high to properly represent orbital changes in the above method, when $1-e$ is very small. The evolution will stop when $e$ is greater than 1.

As shown in Fig. \ref{fig: evolution_wd}, the mass loss is rapidly enhanced after the effects of mass transfer dominate. When the resulting increase in orbital period exceeds the increase in mass loss, $\Delta M$ becomes stable. During this process, the eccentricity keeps growing rapidly until it exceeds 1 and the WD escapes from the SMBH. The period of the last few orbits can vary significantly.

\subsection{Orbital evolution for the main sequence star}
If $\zeta$ of the stripped star is greater than 1/3, such as a main sequence star, then the mass transfer will stabilize and $\langle\dot{\beta}\rangle$ will evolve towards 0. Once this happens, $\langle\dot{\beta}\rangle=0$ will remain. Since the timescale of the mass transfer evolution is larger than that of the orbital parameter evolution, $\dot{M}$ should be able to adjust itself rapidly to balance with the effects of gravitational radiation in the secular evolution. That is, it always satisfies:
\begin{equation}\label{eq: beta_balance}
    \left\langle\frac{\dot{\beta}}{\beta}\right\rangle=\left\langle\frac{\dot{\beta}}{\beta}\right\rangle_{\dot{M}}+\left\langle\frac{\dot{\beta}}{\beta}\right\rangle_{GW}\approx 0.
\end{equation}

\begin{figure}
    \centering
    \includegraphics[width=\linewidth]{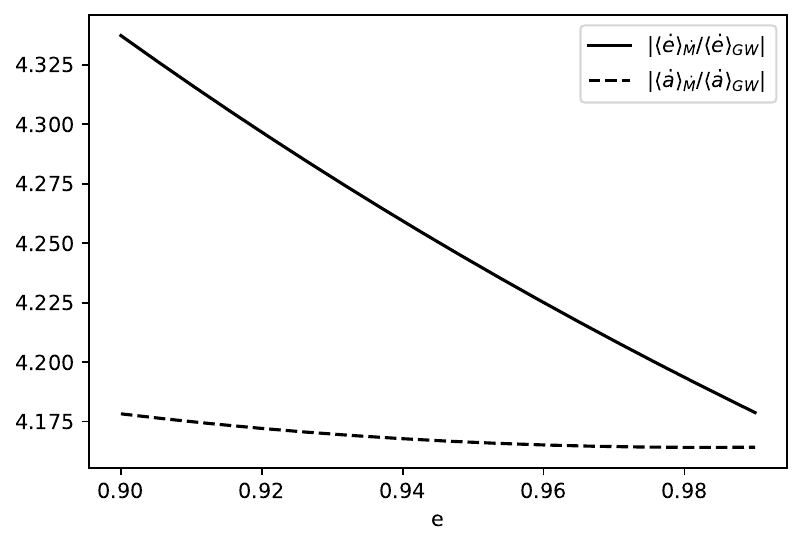}
    \caption{Comparison of $\langle\dot{a}\rangle$ and $\langle\dot{e}\rangle$ caused by the mass transfer and gravitational radiation with $\langle\dot{\beta}\rangle=0$. Here, $\eta$ is taken as 2/3. \label{fig: balance_ms}}
\end{figure}

While these two effects are balanced on $\langle\dot{\beta}\rangle$, they are not balanced on $\langle\dot{a}\rangle$ and $\langle\dot{e}\rangle$. As shown in Fig. \ref{fig: balance_ms}, the effect of the mass transfer is always dominant when $\langle\dot{\beta}\rangle=0$, but it is of the same magnitude as GW. Thus, the evolution of a stripped main sequence star is similar to that of a WD, but on a much longer timescale: the gravitational radiation timescale. This is because $\dot{M}$ is driven by gravitational radiation.

\begin{figure}
    \centering
    \includegraphics[width=\linewidth]{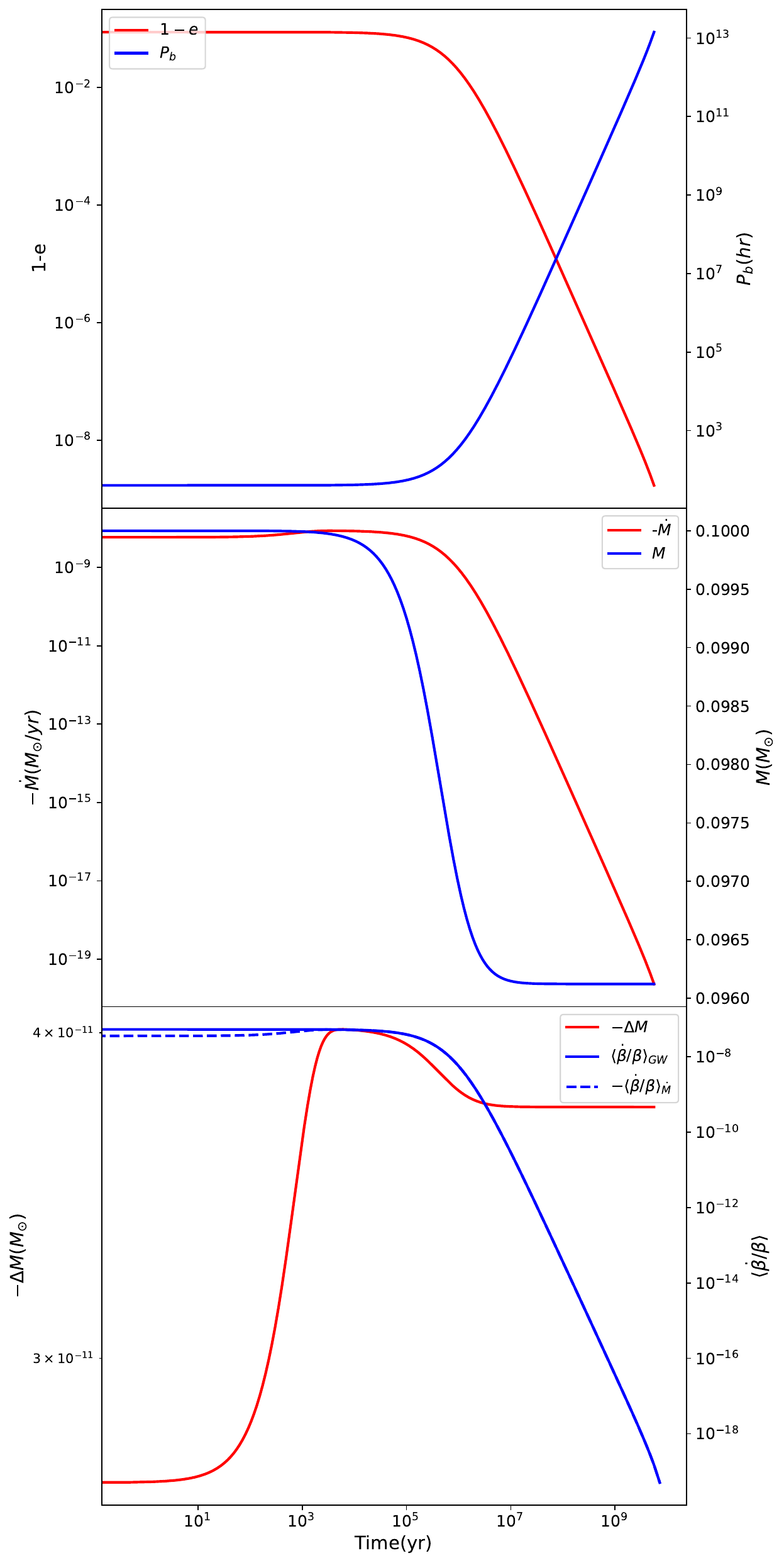}
    \caption{Similar as Fig. \ref{fig: evolution_wd}, but for a main sequence star. The last panel from the top also shows the evolution of $\langle\dot{\beta}\rangle$ of gravitational radiation and mass transfer.\label{fig: evolution_ms}}
\end{figure}

We show an example of this in Fig. \ref{fig: evolution_ms}, where the initial orbital parameters are $M_h=5\times 10^5 M_{\odot}$, $M=0.1 M_{\odot}$, $\beta=0.50004,$ and $P_b=40 \mathrm{~hr}$. The radius is obtained via $R=1.06(M/M_{\odot})^{0.945}R_{\odot}$ \citep{demircan1991stellar}. Then, the eccentricity can be obtained $e=0.912$.

As shown in Fig. \ref{fig: evolution_ms}, the effect of mass transfer is initially weak and $|\dot{M}|$ increase rapidly driven by gravitational radiation and then reaching a balance at $\langle\dot{\beta}\rangle=0$. At this point, the evolution of the orbit is dominated by the effects of mass transfer, with $r_p$ remaining almost constant, while $e$ grows. Since $\langle\dot{\beta}/\beta\rangle_{GW}$ scales as $(1-e)^{3/2}$, $|\dot{M}|$ coupled to it will keep decreasing. Eventually, as in the case of WDs, main sequence stars escape from SMBH, but on much longer timescales. 

There is also a special case where if the initial $|\dot{M}|$ and $e$ are both high, $e$ may evolve to greater than 1 before $|\dot{M}|$ has had time to be reduced to balance state. In addition, in the final stages of evolution, when the orbital period is growing rapidly, there is a possibility that the star may be perturbed by other gravitational sources away from the black hole and, thus, it could escape early.

\subsection{Comparison with the observation of QPEs}

Before the QPEs from GSN 069 disappeared, its peak intensity in 0.4–1 keV decayed dramatically \citep{miniutti2023repeating}. However, the bolometric luminosity by assuming absorbed black-body spectrum was almost constant for each observation, except for the last one, where it was reduced by roughly half \citep{miniutti2023alive}. If the bolometric luminosity is positively correlated with the mass loss of the star at the periastron, such result implies that $\beta$ is constant and then gets smaller.

Assuming the QPEs is caused by the mass transfer via Roche-lobe overflow, the orbiting star should be a WD for GSN 069 \citep{king2020gsn}. As shown in Fig. \ref{fig: evolution_wd}, the mass loss for WD grows rapidly except in the final stage of evolution, when the orbital period also grows rapidly. There was no significant change in the period of QPEs in GSN 069. Thus, a mechanism for stabilizing mass transfer would required to explain the long-term evolution of QPEs and the drag of a gaseous disk may achieve this effect.


\section{Evolution driven by the accretion disk}\label{sec3}
\subsection{Drag force by the accretion disk}
The drag force caused by the accretion disk can be expressed as:
\begin{equation}\label{eq: drag force}
    \mathbf{a}=-\frac{\mathbf{v}_{rel}}{\tau_{F}}.
\end{equation}
Here, $\mathbf{v}_{rel}=\mathbf{v}-\mathbf{v}_{k}$ is the relative velocity of the star with respect to the local gas, where $\mathbf{v}$ is the speed of star and the keplerian velocity, $\mathbf{v}_{k}$, is the velocity of the local gas. The timescale, $\tau_{F}$, of the drag force, which is considered as hydrodynamic force or dynamical friction. The hydrodynamic timescale is \citep{szolgyen2022eccentricity}:
\begin{equation}\label{eq: hydro}
    \tau_H=\frac{2M}{C_D\pi R^2\rho_g v_{rel}}.
\end{equation}
Here, $C_D$ is related to the shape of the star and is equal to 1 for a sphere, while $\rho_g$ is the density of the local gas. The timescale of the dynamical friction is \citep{ostriker1999dynamical}:
\begin{equation}\label{eq: dyn}
    \tau_D=\frac{v_{\mathrm{rel}}^3}{4 \pi G^2 M \rho_{g}F(\mathcal{M})},
\end{equation}
where $F(\mathcal{M})$ is 
\begin{equation}
    F(\mathcal{M})= \begin{cases}\frac{1}{2}\ln \left(\frac{1+\mathcal{M}}{1-\mathcal{M}}\right)-\mathcal{M}, & 0<\mathcal{M}<1, \\ \frac{1}{2} \ln \left(1-\frac{1}{\mathcal{M}^2}\right)+\ln \left(\frac{r}{R}\right), & \mathcal{M}>1.\end{cases}
\end{equation}
Here, $\mathcal{M}=v_{\mathrm{rel}}/c_s$ is the Mach number, where $c_s$ is the local sound speed. Also, $r$ is the characteristic size and we adopted the distance from the central black hole.

The change in the specific angular momentum, $h=\sqrt{G(M_h+M)a(1-e)}$, caused by the drag force can be obtained via: 
\begin{equation}\label{eq: hdot_drag}
    \frac{\dot{h}}{h}=-\frac{1}{h\tau_F}\mathbf{r}\times\mathbf{v}_{rel}=\frac{1}{\tau_F}\left(\frac{\chi}{\sqrt{1+e\cos f}}-1\right),
\end{equation}
where $f$ is the true anomaly, while $\chi$ is 1 for the prograde star and -1 for the retrograde star. And the change in the specific energy $K=-G(M_h+M)/2a$ is: 
\begin{equation}\label{eq: kdot_drag}
    \frac{\dot{K}}{K}=-\frac{1}{K\tau_F}\boldsymbol{v}\cdot\boldsymbol{v}_{rel}=\frac{2}{\tau_F}\left(\frac{(1+2e\cos f+e^2)-\chi(1+e\cos f)^{3/2}}{1-e^2}\right).
\end{equation}
The secular evolution can be obtained by averaging them over an orbital period. Then the secular evolution of orbital parameters are:
\begin{equation}
    \left\langle\frac{\dot{a}}{a}\right\rangle_{disk}=-\left\langle \frac{\dot{K}}{K}\right\rangle,
\end{equation}
\begin{equation}
    \left\langle\dot{e}\right\rangle_{disk}=-\frac{1-e^2}{e}\left(\left\langle\frac{\dot{h}}{h}\right\rangle+\frac{1}{2}\left\langle\frac{\dot{K}}{K}\right\rangle\right),
\end{equation}
and the change in $\beta$ is 
\begin{equation}\label{eq: beta_h_k}
    \left\langle\frac{\dot{\beta}}{\beta}\right\rangle_{disk}=-\frac{1+e}{e}\left\langle\frac{\dot{h}}{h}\right\rangle-\frac{1-e}{2e}\left\langle\frac{\dot{K}}{K}\right\rangle.
\end{equation}
 
When near to the periastron, the velocity of the orbiting star is greater than the Keplerian velocity and it is thus slowed down as it moves away from the periastron. When its velocity becomes lower than the Keplerian velocity, it is therefore accelerated. Since the latter interaction time is longer, the star will likely gain angular momentum from the accretion disk and thus reduce $\beta$.

\begin{figure}
    \centering
    \includegraphics[width=\linewidth]{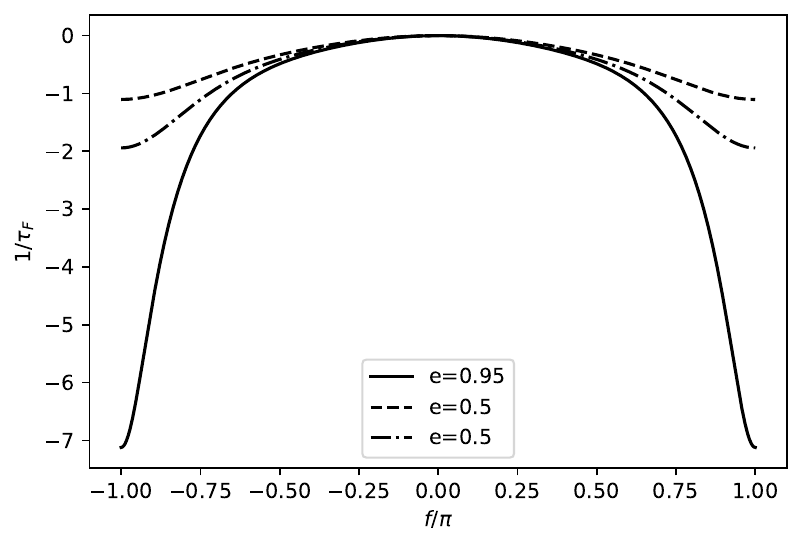}
    \caption{$\dot{\beta}/\beta$ with the true anomaly for a different eccentricity. For all eccentricities, $\dot{\beta}/\beta$ is always negative and equal to zero, while $f=0$.\label{fig: beta_dot_f}}
\end{figure}

Equation \eqref{eq: beta_h_k} can also be used to calculate instantaneous $\dot{\beta}/\beta$ for different values of $f$, as shown in Fig. \ref{fig: beta_dot_f}. Even for a star near the periastron, $\dot{\beta}$ is still positive, so the drag force by the disk may always suppress the mass transfer.

\subsection{Application to GSN 069}
\subsubsection{Stable mass transfer}
\label{sec: stable_mass_transfer}
\begin{figure}
    \centering
    \includegraphics[width=\linewidth]{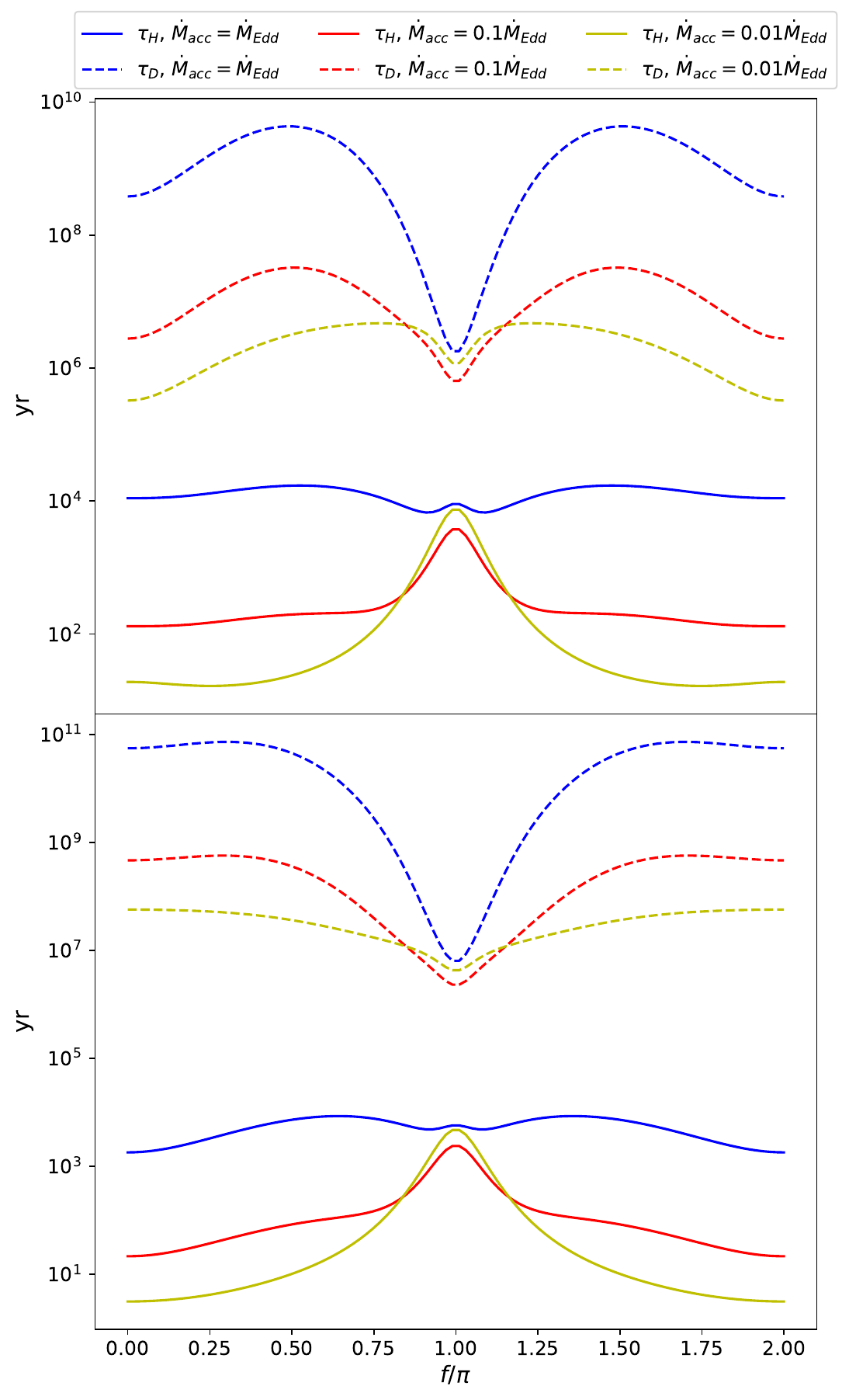}
    \caption{Timescales with the true anomaly for the hydrodynamic force and dynamical friction for a WD with $M=0.21M_{\odot}$, $P_b=9\mathrm{~hr,}$ and $e=0.95$. Upper panel refers to the prograde star and the lower panel to the retrograde star. Solid lines and dashed lines represent the timescales of the hydrodynamic force and dynamical friction, respectively. Different colors represent the different accretion rate of the disk.\label{fig: timescales_wd}}
\end{figure}

We consider an $\alpha$ standard disk with $\alpha=0.1$ and $M_h=5\times 10^5\mathrm{~ M_{\odot}}$. The density and sound speed can be calculated for a given accretion rate, $\dot{M}_{Acc}$, in the unit of Eddington accretion rate, $\dot{M}_{Edd}$ (radiation efficiency $\eta=0.1$). As shown in Fig. \ref{fig: timescales_wd}, the hydrodynamic force is always stronger than the dynamical friction in all orbital phases, thus it dominates the effects from the disk. We only consider the hydrodynamic force in the following calculation.

\begin{figure}
    \centering
    \includegraphics[width=\linewidth]{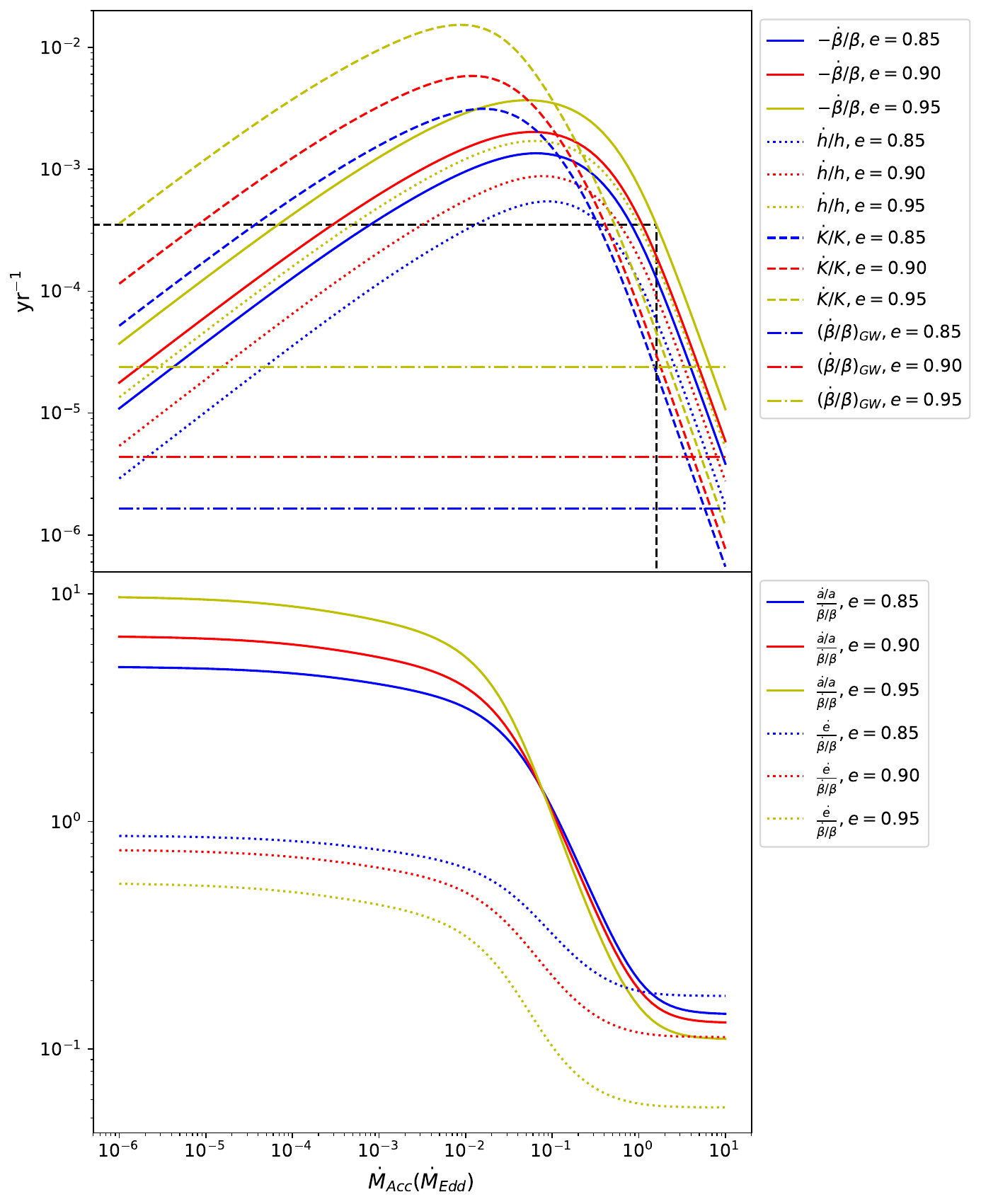}
    \caption{Secular change in the orbital parameters caused by hydrodynamic force for a prograde WD with $M=0.21M_{\odot}$ and $P_b=9\mathrm{~hr}$ under different accretion rate. Top panel shows the change in $\beta$, the specific angular momentum, and the specific orbital energy caused by drag force. It also shows the effect on $\beta$ caused by gravitational radiation for comparison. The dashed black line represents the point with $\dot{M}_{Acc}$ of GSN 069 and $e=0.95$, it predicts $\dot{\beta}/\beta\approx -3.5\times 10^{-4}\mathrm{~yr}^{-1}$(see texts).  Bottom panel represents the change in $a$ and $e$ relative to $\beta$. \label{fig: disk_wd}}
\end{figure}

\begin{figure}
    \centering
    \includegraphics[width=\linewidth]{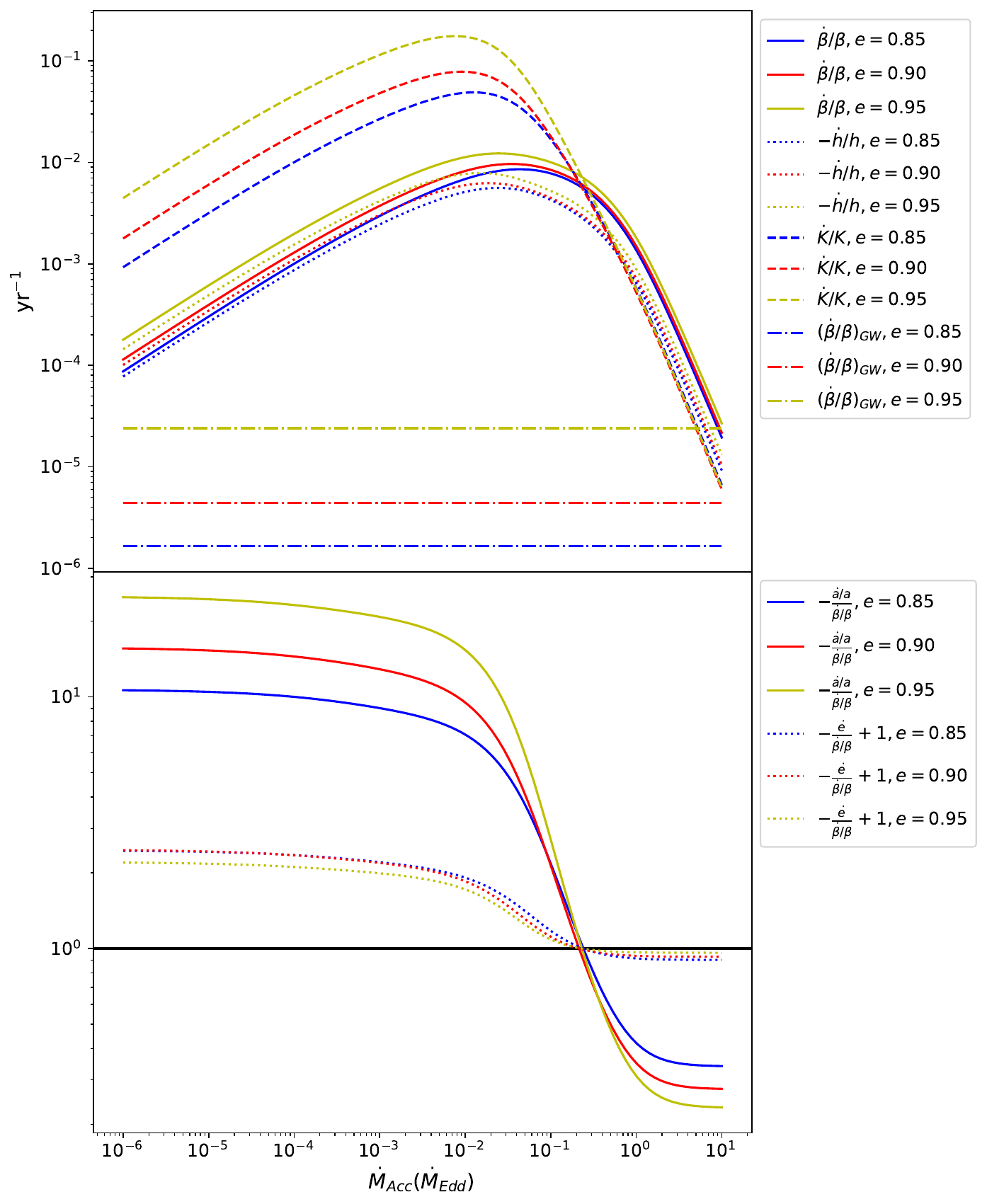}
    \caption{As in Fig. \ref{fig: disk_wd}, but for a retrograde WD. The signs of $\dot{\beta}/\beta$, $\dot{h}/h$, $(\dot{a}/a)/(\dot{\beta}/\beta)$, and $\dot{e}/(\dot{\beta}/\beta)$ are changed. To show a value of $-\dot{e}/(\dot{\beta}/\beta)$ in the logarithmic axis, we add 1 to them because they change sign at a high $\dot{M}_{Acc}$.\label{fig: disk_wd_re}}
\end{figure}

The secular evolution of orbital parameters is shown in Fig. \ref{fig: disk_wd}. For varying $\dot{M}_{Acc}$ and $e$, a prograde WD always gains angular momentum and loses energy from the accretion disk, which decreases $\beta$, $a,$ and $e$. A retrograde WD shows different evolution, as shown in Fig. \ref{fig: disk_wd_re}. The angular momentum is transferred from it to the accretion disk, which makes $\beta$ increase. The energy is still transferred from the WD to the accretion disk. All this makes $\beta$ increase and $a$ continue to decrease, while $e$ continues to decrease at low $\dot{M}_{Acc}$ but increases at high $\dot{M}_{Acc}$.

The different dependence of the parameter evolution on $\dot{M}_{Acc}$ at high and low $\dot{M}_{Acc}$ is attributed to the inner region of the accretion disk being dominated by radiation pressure at high $\dot{M}_{Acc}$ and gas pressure at low $\dot{M}_{Acc}$. Therefore, the disk density decreases with increasing $\dot{M}_{Acc}$ for radiation-pressure dominance resulting in a decrease in drag force, while the disk density increases with increasing $\dot{M}_{Acc}$ for the gas-pressure dominance increasing the drag force.

A prograde WD is preferred for GSN 069. The WD may be captured from a binary system by the Hill mechanism \citep{hills1988hyper,wang2022model}. The accretion disk is likely to be a fallback disk from a TDE, which is possibly a disruption of the common envelope of the binary system \citep{wang2024tidal}. In this scenario, the WD is prograde.

Here, we adopt the results in \cite{miniutti2023repeating}. Assuming a radiation efficiency of $\eta=0.1$, the accreted mass is about $3.7\times 10^{-8} \mathrm{~M_{\odot}}$, corresponding to $\dot{M}/M\approx -1.7 \times 10^{-4} \mathrm{yr}^{-1}$ and $\beta\approx 0.5005$ with $M=0.21 \mathrm{~M_{\odot}}$. The change in $\beta$ induced by mass transfer is $\langle\dot{\beta}/\beta\rangle_{\dot{M}}\approx 1.1\times 10^{-4}\mathrm{~yr}^{-1} $ with $\zeta=-1/3$. The bolometric luminosity of the TDE is about $10^{44}\mathrm{~erg/s}$ corresponding to $\sim 1.6 \dot{M}_{Edd}$. Assuming $e=0.95$, it predicts $\langle\dot{\beta}/\beta\rangle_{disk}\approx -3.5\times 10^{-4}\mathrm{~yr}^{-1}$, as shown in Fig. \ref{fig: disk_wd}. The effects of disk drag and mass transfer on $\beta$ are on the same order of magnitude, suggesting that disk drag has the ability to stabilize mass transfer.

\subsubsection{History and future of evolution}

Early on in the TDE, there is a high accretion rate of the fallback disk, resulting in weak drag force by the disk. The evolution of $\beta$ is dominated by the mass transfer or gravitational wave radiation. Thus, the intensity of QPEs kept increasing during this stage. Overall, QPEs can be observed when the intensity of QPEs is higher than that of the TDE. This is consistent with the fact that QPEs in GSN 069 do not accompany the TDE at the same time, but appear suddenly during the decay of the TDE \citep{miniutti2023repeating}.

As the accretion rate decreases, drag force by the disk begins to dominate $\dot{\beta}$ and, at that point, the intensity of the QPEs begins to decrease. When the TDE rebrightens, the increased accretion rate led to a weaker drag force and, thus, an increase in the mass transfer. Then, when the accretion rate begins to decrease again, the effect of the drag force enhances and the intensity of QPEs will keep decreasing. After switching from radiation-pressure dominance to gas-pressure dominance in the inner region of the disk, the drag force decreases with a decreasing accretion rate. When $\dot{\beta}$ is dominated by gravitational wave radiation, the mass transfer becomes enhanced. When $\dot{\beta}$ is dominated by mass transfer, the unstable mass transfer leads to the WD escaping from the SMBH.

\begin{figure}[hpt]
    \centering
    \includegraphics[width=\linewidth]{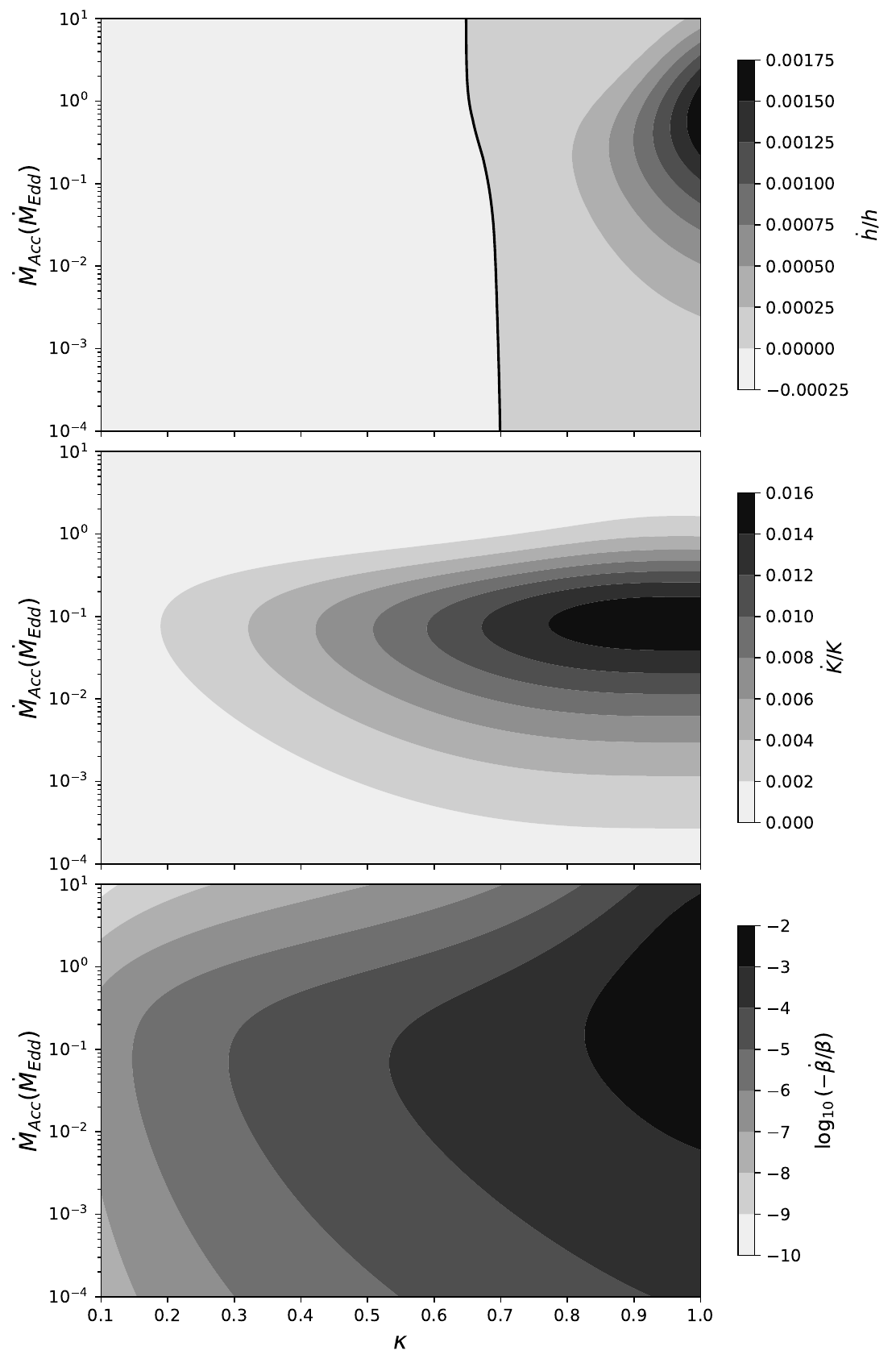}
    \caption{Distribution of secular change in orbital parameters for different $\dot{M}_{Acc}$ and $\kappa$. Top, middle, and bottom panels represent $\dot{h}/h$,$\dot{K}/K,$ and $\dot{\beta}/\beta$, respectively. They are obtained with the drag force acting only when true anomaly $f$ is within $-\kappa\pi$ to $\kappa\pi$. The solid line in the top panel represents $\dot{h}=0$.  \label{fig: f_scale}}
\end{figure}

While the accretion rate decreases, the outer radius of the fallback disk may change \citep{shen2014evolution}. The orbital semi-long axis of the stripped WD may be larger than the outer radius of the disk. In this situation, the drag force by the disk acts only on that side of the periastron. The secular evolution of the orbital parameters can be obtained by considering drag force only at $-\kappa\pi<f<\kappa\pi$ when averaging them over an orbit. Here, $\kappa$ ranges from 0 to 1. As shown in Fig. \ref{fig: f_scale}, the secular evolution of orbital parameters is significantly affected by $\kappa$, especially for $\dot{\beta}/\beta$. When $\kappa<0.5$, the drag force only acts near the periastron; thus, the angular momentum of the stripped WD is transferred to the disk, as expected.

Even though the change in the outer radius of the disk affects the evolution of $\beta$, the tendency for the effect of the drag force by the disk to diminish is constant. Therefore, $\beta$ will start to grow when the effect of gravitational wave radiation exceeds the drag force by the disk. Finally, the stripped WD will escape from SMBH after entering a self-unstable mass transfer, as described in Section \ref{sec2}.

\subsubsection{Secular evolution of GSN 069}

\begin{figure}[hpt]
    \centering
    \includegraphics[width=\linewidth]{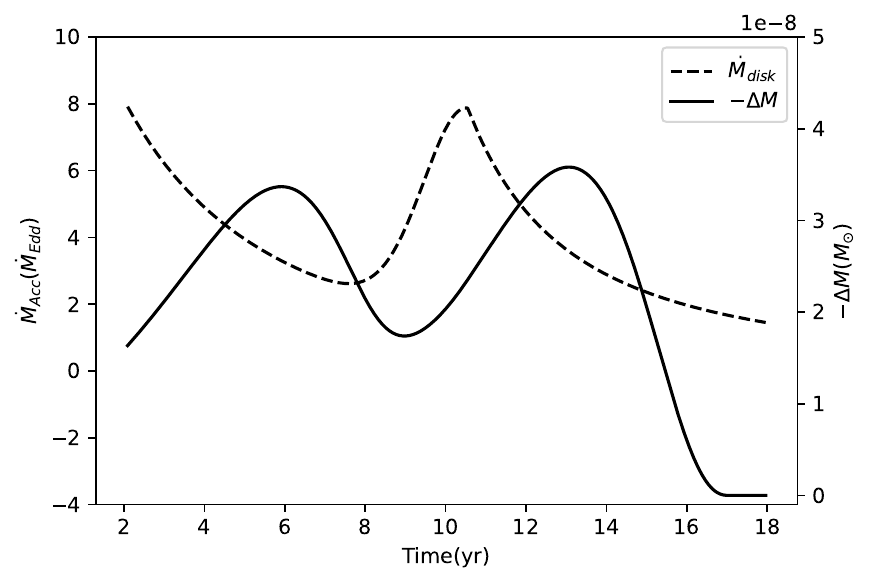}
    \caption{Possible evolution of mass transfer with the evolution of the accretion rate of the accretion disk. Solid line and dashed lines represent the evolution of the mass transfer at the periastron and the accretion rate of the disk in GSN 069, respectively. See text for details of the calculations.\label{fig: evolution with disk}}
\end{figure}

Due to the evolution on the accretion rate of the fallback disk in GSN 069, the secular evolution on mass transfer may experience complex changes. Figure \ref{fig: evolution with disk} shows an example with possible parameters of GSN 069. Following the formula in \cite{miniutti2023repeating}, the accretion rate of each peak for the TDE is described by:
\begin{equation}
    \dot{M}=\dot{M}_{\text {peak }} \times \begin{cases}e^{-\left(t-t_{\text {peak }}\right)^2 / 2 \sigma^2} & t \leq t_{\text {peak }} \\ {\left[\left(t-t_{\text {peak }}+t_0\right) / t_0\right]^n} & t>t_{\text {peak }}\end{cases},
\end{equation}
where the parameters can be obtained by fitting the light curve \citep{miniutti2023repeating}. For the first peak, $t_0=7.6$ years and $t_{\text {peak }}=600$ days, only the decay phase is calculated. For the second peak, $t_0=4.8$ year, $t_{\text {peak }}=3850$ days and $\sigma=400$ days. And $n$ is -9/4 for each peak, $\dot{M}_{\text {peak }}$ is 9 $\dot{M}_{Edd}$ for the first peak and 6.3 $\dot{M}_{Edd}$ for the second peak. The orbital parameters are same as those given in  Section \ref{sec: stable_mass_transfer}, while the time is set to $7$ years. As a result, the past and future evolution can be obtained using the same method presented in Section \ref{sec: orbital_evolution_for_the_wd}.

It should be stressed that the results in Fig. \ref{fig: evolution with disk} are very sensitive to the magnitude of parameters, mainly $\beta$ and $\dot{M}_{\text {peak }}$. The former affects the magnitude of $\dot{M}$ as it enlarges by a factor of $1/(\beta-\beta_0)$ compared to the $\dot{\beta}$, as shown in Eq. 3. The latter, on the other hand, affects the magnitude of the disk drag, thus changing the moment of transition when $\beta$ changes from increasing to decreasing.

We do not show the evolution of the light curve directly, because the radiative processes are not considered here. If we assume a positive correlation between the bolometric luminosity and the accretion mass, the disappearance of the QPEs in the early stages of the TDE and the lack of increase of its luminosity (as well as its fading) can be explained. That is to say, the early accreted mass increases rapidly and then starts to decrease as the disk drag increases.

However, the subsequent evolution is somewhat difficult to explain. Although the decrease in the accreted mass when the TDE flares again seems to explain the disappearance of the QPEs, as pointed out by \cite{miniutti2023alive}, the spectrum of QPEs need to be colder than it of the TDE to be undetectable -- and not just less luminous than the TDE. In addition, the second flare of the two QPEs that reappeared brightened (as expected from Fig. \ref{fig: evolution with disk}), however, the first flare did not brighten and could not be explained by our model.

These variations may be related to the radiation processes within the QPEs. \cite{lu2023quasi} suggested that the QPEs is produced by the circularized shock of the accreted matter moving inside the disk. The peak temperature decreases during the second outburst of the TDE due to the decrease in disk density. This may explain the disappearance of QPEs. Thus, the evolution of QPEs is also directly related to the evolution of accretion disk. The secular evolution on the light curve and spectrum of QPEs requires a further analysis combined with radiative processes in the future.

For the accretion model of the QPEs, the long-short behavior on the intensity and recurrence time is difficult to explain. \cite{king2022quasi} proposed that the short accretion timescale may cause the WD to fail to maintain hydrostatic equilibrium and oscillate in radius, which may explain this feature of QPEs. Such oscillations may be perturbed when the density of the surrounding gas changes, which may explain the disappearance of such features in QPEs before the second outburst of the TDE and the weaker QPE after this outburst that did not become enhanced, as we expected. The evolution of such short timescales is beyond the scope of this paper, which focuses on long-term evolution. 

\section{Discussion} \label{sec4}
\subsection{Unconservative mass transfer}
It is also possible for material to be lost through the Lagrangian point L2 and this is more likely to happen for systems with high eccentricity \citep{sepinsky2007equipotential}. This lost material may form a circumbinary binary disk, which may significantly affect the orbital evolution\citep{dermine2013eccentricity}. This scenario is not discussed here,  in favor of focusing on the case where this material leaves the binary system altogether. 

Differentiating the total angular momentum gives: 
\begin{equation}\label{eq: dJ}
    \frac{\dot{J}}{J}=\frac{\dot{M_h}}{M_h}+\frac{\dot{M}}{M}-\frac{\dot{M_t}}{2M_t}+\frac{1}{2}\frac{\dot{e}}{1+e}+\frac{1}{2}\frac{\dot{r}_p}{r_p}.
\end{equation}
Assuming the mass transfer occurs only instantaneously at the periastron, then $\dot{r}_p=0$. The fraction of mass that is accreted to the central black hole is defined as $\gamma=-\dot{M}_h/\dot{M}$. The mass lost by the system also carries away angular momentum. Based on the assumption that the ratio of its specific angular momentum to that of the stripped star is $\lambda$, the change in the total angular momentum is $\dot{J}/J=\lambda\dot{M}_t/qM_t$. Then Eq. \eqref{eq: dJ} becomes:
\begin{equation}\label{eq: edot_mdot_l2}
    \langle\dot{e}\rangle_{\dot{M}}=-\mathcal{F}(\gamma,q,\lambda)(1+e)\frac{\dot{M}}{M},
\end{equation}
where 
\begin{equation}
    \mathcal{F}(\gamma,q,\lambda)=\frac{2+2(\gamma-1)\lambda+(1-\gamma)q-2\gamma q^2}{1+q}.
\end{equation}
By $\dot{r}_p=0$, we have:
\begin{equation}\label{eq: adot_mdot_l2}
    \left\langle\frac{\dot{a}}{a}\right\rangle_{\dot{M}}=\frac{\langle\dot{e}\rangle_{\dot{M}}}{1-e}=-\mathcal{F}(\gamma,q,\lambda)\frac{1+e}{1-e}\frac{\dot{M}}{M}.
\end{equation}
When $\gamma=1$, these are reduced to Eq. \eqref{eq: edot_mdot} and \eqref{eq: adot_mdot}. 

Assuming that half the mass is lost from L2 and their specific angular momentum is the same as that of the stripped star, i.e., $\gamma=0.5$ and $\lambda = 1$, Eq. \eqref{eq: edot_mdot_l2} becomes
\begin{equation}
    \left\langle\dot{e}\right\rangle_{\dot{M}}=-\frac{1+0.5q-q^2}{1+q}\frac{\dot{M}}{M}(1+e)\approx-(1+e)\frac{\dot{M}}{M}\left(1-\frac{q}{2}\right).
\end{equation}
This result is about half of what it is at $\gamma=1$, as is $\langle\dot{a}/a\rangle_{\dot{M}}$. The change in $\beta$ now is: 
\begin{equation}
    \left\langle\frac{\dot{\beta}}{\beta}\right\rangle_{\dot{M}}\approx \frac{\dot{M}}{M}\left(\zeta-\frac{1}{3}-\frac{q}{2}\right),
\end{equation}
which is approximately same as Eq. \eqref{eq: beta_dot_mdot} for $q\ll 1$. 

In this scenario, the effect of mass transfer is about twice the effect of the gravitational radiation in the balance state by comparing it with the case of $\gamma=1,$ as shown in Fig. \ref{fig: balance_ms}. Thus, the effect of mass transfer is still dominant on the orbital evolution in the balance state for the main sequence star. 

On the other hand, in the presence of a large-scale disk, material passing through L2 is injected into the disk and then accreted by the black hole. Therefore, the loss of matter through L2 does not affect the previous results.

\subsection{Extreme mass ratio inspirals}
Extreme mass-ratio inspirals are potential gravitational wave sources and they are expected to form when compact objects gradually inspiral via gravitational wave radiation after being captured on a highly eccentricity orbit \citep{amaro2018relativistic}. Our results show that if the mass transfer is triggered during this process, the inspiral will be terminated and will thus fail to become a gravitational wave source.

\subsection{Considering the other QPEs}
GSN 069 is not the only galaxy found to have QPEs \citep{giustini2020x,song2020possible,arcodia2021x,chakraborty2021possible}, but is the only one with both TDEs and QPEs \citep{shu2018long,sheng2021evidence,miniutti2023repeating}. The orbiting WD is likely captured from a binary via the Hills mechanism \citep{hills1988hyper,wang2022model}, while the TDE possibly originates from a disruption of a common envelope \citep{wang2024tidal}. In this situation, the fallback disk is coplanar with the orbit of the common envelope, which is almost coplanar with the orbit of the captured WD. Thus, the orbiting star moves inside the disk for GSN 069.

As for the other QPE sources, the situation may be different. The pre-existing disk may not be present or the pre-existing disk may be inclined to the orbit of the stripped star. In the former case, the orbital evolution is the same as discussed in Section \ref{sec2}. In the latter case, the effect of the disk on the orbit will be different from that discussed in Section \ref{sec3}. For example, \cite{linial2024period} found the hydrodynamic drag force received by the star as it passes through the inclined disk can make the orbital period decrease. 

The TDE in GSN 069 makes this system unique. On the one hand, it may be the only source whose accretion disk is coplanar with the orbit of the stripped star. On the other hand, the accretion rate of its disk is dominated by the TDE and, therefore, it evolves rapidly. These two points lead to a complicated orbital evolution, making the QPEs in GSN 069 an excellent source for studying the interaction between the disk and the object moving inside it.

\section{Summary} \label{sec5}

By accounting for the effect of the mass transfer, the evolutionary path of the stripped star is different from the evolutionary only with gravitational wave radiation. The criterion of the unstable mass transfer is changed and the fate of the stripped star is to escape the SMBH, rather than be disrupted by it. In this case, however, mass transfer remains unstable.

The TDE in GSN 069 may be a disruption of the common envelope before the WD was captured, so the orbiting WD is likely moving inside the fallback disk. The hydrodynamic drag force by the disk can effectively reduce $\beta$ and stabilize the mass transfer, which may explain the observed long-term evolution in the intensity of QPEs. After the TDE is re-brightened, the mass transfer increases due to the weakening of the drag force caused by the reduced density of the inner zone of the accretion disk. As the accretion rate of the fallback disk gradually decreases, the drag force by the disk will decrease until it is negligible. Then the mass transfer will start to be unstable  the WD eventually escapes the SMBH. 

The spectrum evolution of QPEsis also directly related to the evolution of the accretion disk. Future works will require further quantitative analyses of the evolution of mass transfer in combination with specific radiative processes. In addition, continued observations of GSN 096 will help us to understand the interaction of the disk with the star inside it and to test our model.

\begin{acknowledgements}
    Thanks Mengye Wang for helpful discussion. This work was supported by the National Key Research and Development Program of China (No. 2020YFC2201400) and National SKA Program of China  (2020SKA0120300).
\end{acknowledgements}

\bibliography{ref} 
\bibliographystyle{aa}
\end{document}